\documentclass[12pt,a4paper]{article}

\DeclareFontFamily{OT1}{rsfs}{}
\DeclareFontShape{OT1}{rsfs}{m}{n}{ <-7> rsfs5 <7-10> rsfs7 <10->
rsfs10}{} \DeclareMathAlphabet{\mycal}{OT1}{rsfs}{m}{n}

\newcommand{\be}{\begin{equation}}
\newcommand{\ee}{\end{equation}}
\newcommand{\beq}{\begin{eqnarray}}
\newcommand{\eeq}{\end{eqnarray}}

\usepackage{amsmath,amsthm,amsfonts}
\input amssym.def
\newsymbol\square 1003

\begin{document}

\centerline{\bf  NEW DESCRIPTION OF SELF-DUAL METRICS}

\medskip

\centerline{J. Tafel}

\noindent
Institute of Theoretical Physics, University of Warsaw,
Ho\.za 69, 00-681 Warsaw, Poland, email: tafel@fuw.edu.pl

\bigskip

\noindent
{\bf Abstract}. We show that Pleba\'nski's equation for self-dual metrics is equivalent to a pair of equations describing canonical transformations in 2-dimensional phase spaces. Examples of linearizations of these equations are given.

\bigskip


\null

\section{Introduction}
Pleba\'nski \cite{Pl} showed that complex metrics with the self-dual Riemann tensor \cite{P}, to be refferred to as self-dual metrics, can be described in terms of one function $\Omega$ satisfying so-called Pleba\'nski's first equation
\be
\Omega_{,q\tilde q}\Omega_{,p\tilde p}-\Omega_{,q\tilde p}\Omega_{,p\tilde q}=1.\label{1}
\ee
 Given $\Omega$ the corresponding metric reads
\be
g= \Omega_{,q\tilde q}dq d\tilde q+\Omega_{,p\tilde p}dp d\tilde p+\Omega_{,q\tilde p}dq d\tilde p+\Omega_{,p\tilde q}dp d\tilde q.\label{2}
\ee
He also proposed an equivalent equation known as Pleba\'nski's second equation.
At present there are several other descriptions of self-dual metrics  \cite{A,C,G,H} (note that an erroronous sign in Husain's equation was corrected and an equivalence of the latter equation to (\ref{1}) was proved \cite{T}). Much attention was payed to possible complete integrability of the self-duality conditions (see \cite{MW} and references therein) .

In  this paper we represent self-dual metrics in the form
\be 
g=\{u,v\}_{q\tilde q}dq d \tilde q+\{u,v\}_{p\tilde p}dp d \tilde p+\{u,v\}_{q\tilde p}dq d \tilde p+\{u,v\}_{p\tilde q}dp d \tilde q,\label{3}
\ee
where the Poisson  brackets are taken with respect to indicated coordinates,\linebreak
 e.g. $\{u,v\}_{q\tilde q}=u_{,q}v_{,\tilde q}-u_{,\tilde q}v_{,q}$.
 Functions $u$ and $v$ are subject to the equations
\be 
\{u,v\}_{qp}=1\label{4}
\ee
\be
\{u,v\}_{\tilde q\tilde p}=1\ .\label{4a}
\ee
Thus, $u$ and $v$ are related to $q$, $p$ and to $\tilde q$, $\tilde p$ by canonical transformations. Solving (\ref{4}) and (\ref{4a}) requires correlating these transformations. In section 3 we show that these equations can be linearized under some assumptions on functions generating the canonical transformations.

It is convenient to denote coordinates $q$, $p$ by $z^A$, $A=1,2$, and coordinates $\tilde q$, 
$\tilde p$ by $\tilde z^A$. Partial derivatives of any function $f$ with respect to these coordinates will be denoted by $f_{,A}$ and $f_{,\tilde A}$, respectively. Instead of $\{u,v\}_{z^A\tilde z ^B}$ we will write  $\{u,v\}_{A\tilde B}$.  In this notation metric (\ref{2}) reads
\be
g=\Omega_{,A\tilde B}dz^Ad\tilde z^{B}\label{4b}
\ee
and metric (\ref{3}) reads
\be
g=\{u,v\}_{A\tilde B}dz^Ad\tilde z^{B}.\label{5}
\ee
 
\section{Equivalence of formulations}

Let $u$ and $v$ satisfy equations (\ref{4}) and (\ref{4a}). Then 
\be
\{u,v\}_{A[\tilde B,\tilde C]}=\{u,v\}_{\tilde B [A,C]}=0,\label{5a}
\ee
hence there are functions $\Omega_A$ such that 
\be
\{u,v\}_{A\tilde B}=\Omega_{A,\tilde B}\label{5b}
\ee
and
 $\Omega_{[A,C]}$ do not depend on coordinates $\tilde z^{A}$. One can shift $\Omega_A$ by functions of $z^A$ to achieve $\Omega_{[A,C]}=0$. From this property it follows that there is  a function $\Omega$ such that $\Omega_A=\Omega_{,A}$. Thus, 
\be
\{u,v\}_{A\tilde B}=\Omega_{,A\tilde B}\label{6}
\ee
and metric (\ref{5}) takes the form (\ref{4b}). Substituting (\ref{4}), (\ref{4a}) and (\ref{6}) into the identity 
\be
\{u,v\}_{q\tilde q}\{u,v\}_{p\tilde p}-\{u,v\}_{q\tilde p}\{u,v\}_{p\tilde q}=\{u,v\}_{qp}\{u,v\}_{\tilde q\tilde p}\label{7}
\ee
shows that $\Omega$ satisfies  Pleba\'nski's  equation (\ref{1}).

To prove that the reverse transformation  exists is less straightforward. Let us use the antisymmetric tensors $\epsilon^{AB}$ and $\epsilon^{\tilde A\tilde B}$ to raise indices in a standard way, e.g. $p^{,A}=\epsilon^{AB} p_{,B}$. This notation allows to write equations (\ref{4}) and (\ref{4a}) in the form
\be
v^{,A}u_{,A}=1\label{8}
\ee
\be
v^{,\tilde A}u_{,\tilde A}=1.\label{9}
\ee
Assume that a function $\Omega$ satisfies the equation
\be
\Omega_{,A\tilde B}\Omega_{,C}^{\ \ \tilde B}=\epsilon_{AC},\label{10}
\ee
which is equivalent to (\ref{1}). Our aim is to prove the existence of solutions  $u,v$ of 
equations (\ref{6}), (\ref{8}) and (\ref{9}).

Taking contractions of  (\ref{6}) with $v^{,A}$ or $u^{,A}$ and using (\ref{8}) we can replace equations (\ref{6}) by
\be
v_{,\tilde B}=
v^{,A}\Omega_{,A\tilde B}\label{11}
\ee
and
\be
u_{,\tilde B}=
u^{,A}\Omega_{,A\tilde B}.\label{12}
\ee
It follows from (\ref{10})-(\ref{12}) that
\be
v^{,\tilde B}u_{,\tilde B}=v^{,A}u^{,C}(\Omega_{,A}^{\ \ \tilde B}\Omega_{,C\tilde B})=v^Au_A\ .
\label{12a}
\ee
Thus, equation (\ref{9}) is satisfied if  equations (\ref{8}) and (\ref{10})-(\ref{12}) are satisfied.

Let us write equations (\ref{11}) and (\ref{12}) as
\be
D_{\tilde B}v=0\label{13}
\ee
\be
D_{\tilde B}u=0,\label{14}
\ee
where operators $D_{\tilde B}$ are given by 
\be
D_{\tilde B}=\partial_{\tilde B}+\Omega^{,A}_{\ \ \tilde B}\partial_A.\label{15}
\ee
Vectors $D_{\tilde B}$ are part of a null basis for metric (\ref{2}). In view of twistor constructions related to self-dual metrics \cite{PR} it is not suprising that $D_{\tilde B}$ commute provided equation (\ref{10}) is satisfied,
\be
[\partial_{\tilde B}+\Omega^{,A}_{\ \ \tilde B}\partial_A,\partial_{\tilde C}+\Omega^{,D}_{\ \ \tilde C}\partial_D]=2\Omega^{,A}_{\ \ [\tilde B}\Omega^{,D}_{\ \ \tilde C],A}\partial_D=
2(\Omega^{,A}_{\ \ [\tilde B}\Omega^{,D}_{\ \ \tilde C]})_{,A}\partial_D=0\ .\label{15a}
\ee
 Let $v$ be a solution of (\ref{13}). It remains to solve equations (\ref{8}) and (\ref{14}), which form a linear system for a function $u$. The system is integrable since all the operators $D_{\tilde B}$, $v^{,A}\partial_{A}$ commute due to  (\ref{15a}) and the equality
\be
[\partial_{\tilde B}+\Omega^{,C}_{\ \ \tilde B}\partial_C,v^{,A}\partial_{A}]=(v_{,\tilde B}+\Omega^{,C}_{\ \ \tilde B}v_{,C})^{,A}\partial_{A}=0\label{15b}
\ee
forced by (\ref{13}).
Thus, equations (\ref{6}), (\ref{8}), (\ref{9}) have solutions and we obtain the following theorem.

\newtheorem{tw}{Theorem}
\begin{tw}
All complex self-dual metrics can be locally represented in the form (\ref{3}), where $u$ and $v$ are subject to equations (\ref{4}) and (\ref{4a}).
\end{tw}

Note that if $u$, $v$ and all the coordinates are real, then metric (\ref{3}) is real with the signature $++--$. The same signature is obtained when $u$ and $iv$ are real and 
\be
\tilde q=\epsilon \bar q\ ,\ \tilde p=-\epsilon\bar p,\   \epsilon=\pm 1\ , \label{19}
\ee
where the bar denotes the complex conjugate.
Simple conditions of this kind are not available for the Euclidean signature of $g$. This difficulty appears also in other approaches to self-duality \cite{G,H}.

Equations (\ref{4}) and (\ref{4a}) possess  point symmetries which are easy to obtain if we denote $u$, $v$ by $f^A$ and write equations (\ref{4}), (\ref{4a}) as
\be
det\big (\frac{\partial f^A}{\partial z^B}\big )=det\big (\frac{\partial f^A}{\partial \tilde z^B}\big )=1.\label{19a}
\ee
Let new variables $f'^A$ be functions of $f^B$, new coordinates $z'^A$ be functions of $z^B$ and $\tilde z'^A$ be functions of  $\tilde z^B$. This transformation preserves equations  (\ref{19a}) provided
\be
det\big (\frac{\partial f'^A}{\partial f^B}\big )=det\big (\frac{\partial z'^A}{\partial z^B}\big )=det\big (\frac{\partial \tilde z'^A}{\partial \tilde z^B}\big )=c,\label{19b}
\ee
where $c$ is a constant. In terms of Poisson brackets equations (\ref{19b}) read 
\be
\{u',v'\}_{uv}=\{q',p'\}_{qp}=\{\tilde q',\tilde p'\}_{\tilde q\tilde p}=c.\label{20}
\ee
 There are also natural discrete symmetries given by an appropriate  interchange of variables. All above  transformations preserve the metric modulo a constant factor.

\section{Reductions to linear equations}

In terms of a generating function $S(v,q,\tilde q,\tilde p)$, where $\tilde q$, $\tilde p$ appear as parameters, solutions $u$, $v$ of (\ref{4}) are given implicitly by
\be
p=S_{,q}\ ,\ \ u=S_{,v}\ .\label{21}
\ee
Similarly, equation (\ref{4a}) can be formally solved by means of a generating function $\tilde 
S(v,\tilde q,q,p)$,
\be
\tilde p=\tilde S_{,\tilde q}\ ,\ \ u=\tilde S_{,v}\ .\label{22}
\ee
Functions $S$ and $\tilde S$ should be correlated in such a way that solutions $u$, $v$ obtained from (\ref{21}) or (\ref{22}) coincide. 

Assume that $S$ and $\tilde S$ are linear in $\tilde p$ and $p$, respectively,
\begin{align}
&S=S'(v,q,\tilde q)+\tilde pa(v,q,\tilde q)\ ,\ \ \tilde S=\tilde S'(v,q,\tilde q)+p\tilde a(v,q,\tilde q)\ . \label{24}
\end{align}
Substituting (\ref{24}) into (\ref{21}) and (\ref{22}) leads to the following conditions guaranteeing the uniqueness of $u$ and $v$
\be
\tilde a_{,\tilde q}=a_{,q}^{\ -1}\ ,\ \ \tilde a_{,v}=a_{,v}a_{,q}^{\ -1}\label{25}
\ee
\be
S'_{,q}+a_{,q}\tilde S'_{,\tilde q}=0\ ,\ \ S'_{,v}-\tilde S'_{,v}+a_{,v}\tilde S'_{,\tilde q}=0\ .
\label{26}
\ee
Equations (\ref{25}) form a nonlinear system for the functions $a$ and $\tilde a$. Given $a$ and $\tilde a$  it remains to solve linear equations (\ref{26}) for functions $S'$ and $\tilde S'$.

Equations (\ref{25}) are integrable with respect to $\tilde a$ provided $a$ satisfies the following quasilinear equation
\be
a_{,vq}+a_{,q}a_{,v\tilde q}-a_{,v}a_{,q\tilde q}=0.\label{27}
\ee
Simple solutions of (\ref{27}) can be found by assuming  that  one of the second derivatives of $a$ vanishes. For instance, for $a_{,vq}=0$  one obtains $a=\tilde q(v+q)$. For this function $a$ equations  (\ref{26}) yield
\be
S=F_{,q}+\tilde p\tilde q(v+q),\label{28}
\ee
where function $F(v,q,\tilde q)$ is subject to the linear equation
\be
F_{,vq}+F_{,v\tilde q}-F_{,q\tilde q}=0.\label{29}
\ee
In this case the corresponding metric (\ref{3}) possesses the translational Killing vector $\partial_p-\partial_{\tilde p}$. For this reason it must belong to the class of generalized Hawking metrics \cite{H}. 

If $a_{,v\tilde q}=0$ then  the generating function $S$ can be put into the form
 \be
S=F_{,v}-e^qF_{,\tilde q}+\tilde p(q\tilde q+e^qv+c(q)).\label{30}
\ee
Here $c$ is an arbitrary function of $q$, with its derivative denoted by $\dot c$,
 and  $F$ has to satisfy the linear equation
\be
F_{,vq}+(\tilde q+e^qv+\dot c)F_{,v\tilde q}-e^qF_{,q\tilde q}-e^qF_{,\tilde q}=0.\label{31}
\ee
If $a_{,q\tilde q}=0$ one obtains 
\be
S=F_{,\tilde q}+\tilde p(v\tilde q+e^{-v}q+c(v))\label{32}
\ee
and the following linear equation for $F$
\be
e^{-v}F_{,v\tilde q}+(e^{-v}q-\tilde q-\dot c)F_{,q\tilde q}+F_{,vq}+F_{,q}=0.\label{33}
\ee
 Given a solution $F$ of equation (\ref{31}) or (\ref{33}) one can find functions $u$, $v$ from relations (\ref{21}) and then construct the corresponding self-dual metric (\ref{3}). Up to the best knowledge of the author the reduction of the Pleba\'nski equation, for some ansatz,  to the linear equations (\ref{31}) or (\ref{33}) is new.

\section{Conclusions}
We have shown that all self-dual metrics can be represented in the form (\ref{3}), where functions $u$, $v$ have to satisfy equations (\ref{4}) and (\ref{4a}) defining canonical transformations in a 2-dimensional phase space. 

The new formulation creates new possibilities of construction of self-dual metrics. Equations (\ref{4}), (\ref{4a}) are equivalent to linear ones (\ref{29}), (\ref{31}) or (\ref{33}) under the assumption that functions $u$, $v$ are given by  (\ref{21}) and $S$ takes the form (\ref{28}), (\ref{30}) or (\ref{32}). An open problem is whether these linear reductions admit interesting solutions other than  gravitational instantons of Hawking \cite{Haw}.  

In our opinion the new formulation can be also useful  as a new tool in classification of  completely integrable equations.  It is well known that many of these equations can be considered as reductions of the self-dual Yang-Mills equations \cite{W}  or  the self-dual Einstein equations \cite{DMW} (see also \cite{MW}). Perhaps, due to a new form of self-duality conditions, one can obtain in this way further completely integrable equations.

\bigskip
\noindent {\bf Acknowledgements}. 
This work was partially supported by the Polish  Committee for Scientific Research (grant 1 P03B 075 29).


\begin{thebibliography}{99}
\bibitem{Pl} Pleba\'nski J 1975 \emph{J. Math. Phys.} {\bf 16} 2395

\bibitem{P} Penrose R 1976  \emph{Gen. Rel. Grav.} {\bf 7} 31

\bibitem{A} Ashtekar A, Jacobson T and Smolin L 1988 \emph{Commun. Math. Phys.} {\bf 115} 631

\bibitem{C} Chakravarty S, Mason L J and Newman E T 1991 \emph{J. Math. Phys.} {\bf 32} 1458

\bibitem{G} Grant J D E 1993 \emph{Phys. Rev.} {\bf D48} 2606

\bibitem{H} Husain V 1994 \emph{Class. Quantum Grav.} {\bf 11} 927

\bibitem{T} Jakimowicz M and Tafel J 2005 \emph{work in preparation}

\bibitem{MW} Mason L J and Woodhoue N M J  1996 \emph{Integrability, self-duality and twistor theory} (Clarendon vress, Oxford)

\bibitem{PR} Penrose R and  Rindler W 1986  \emph{Spinors and spacetime}, vol. II (Cambridge Univerity Press, Cambridge)

\bibitem{Haw} Hawking S W 1977 \emph{Phys. Let.} {\bf 60A} 81


\bibitem{W} Ward R S 1985  \emph{Phil. Trans. R. Soc.} {\bf A315} 451

\bibitem{DMW} Dunajski M, Mason L J  and Woodhouse N M J 1998 \emph{J. Phys. A: Math. Gen.} {\bf 31} 6019

\end{thebibliography}
\end{document}